\documentclass[twocolumn,showpacs,amsfonts,amsmath,amssymb,prb]{revtex4}
\usepackage{graphicx}
\usepackage{amsmath}
\usepackage{amssymb}
\newcommand{\rmi}{\text{\fontfamily{cmr}\selectfont i}}
\newcommand{\rme}{\text{\fontfamily{cmr}\selectfont e}}
\newcommand{\arc}{\text{\fontfamily{cmr}\selectfont arc\!}}

\newcommand{\rmV}{\text{\fontfamily{cmr}\selectfont V}}
\begin{document}
\title{Continuum of extended states in the spectrum of a one-dimensional random potential}%
\author{Alberto Rodr\'{\i}guez}%
   \email{argon@usal.es}%
\author{Jose M. Cerver\'o}%
    \affiliation{F\'{\i}sica Te\'orica. Facultad de %
    Ciencias. Universidad de Salamanca. 37008 Salamanca. Spain}%
\begin{abstract}%
    We describe a one-dimensional disordered system, based on the P\"oschl-Teller potential, 
    that exhibits a continuum of extended states which is independent of the random or
    correlated character of the sequence and of the length of the
    system. The delocalization of the electronic states occurs in the whole
    positive spectrum where the system shows a perfect transmission. 
\end{abstract}
\maketitle
Since the work of Anderson \cite{And109_58} localization has been considered
as a central tenet of the theory of disordered systems. Its significance as
an unavoidable consequence of the presence of disorder in the systems was
later enhanced by the results of the Scaling Theory \cite{AbrAnd42_79} that
originially predicted the localization of all electronic states for any degree of
disorder in 2-D and 1-D structures while the existence of a metal to
insulator transition (MIT)  was permitted in 3-D systems
\cite{LeeRam57_85}. It was subsequently shown that in several
configurations of 2-D electron and hole systems a MIT could occur as a
function of the density of carriers \cite{Kra50_94} or as a result of an applied magnetic
field \cite{2dmag}. Such experimental observations meant the reopening of
the localization problem and the Scaling Theory was enlarged to describe 
these new results. In one-dimensional systems the appearance of disordered models
exhibiting short-range and long-range correlations also showed that extended states can
exist in the spectrum of a 1-D disordered structure. Short-range
correlations lead to the emergence of isolated extended states in the
thermodynamic limit, that constitute a zero measure set of the spectrum \cite{shortrange,SanMac49_94,CerRodEPJB}, while
long-range correlations give rise to the appearance of a phase of
apparently extended states and therefore a qualitative MIT in 1-D
\cite{longrange}. Correlations can alter the random character of
the structures and improve noticeably their transport properties as it has
been experimentally verified in different systems such as semiconductor
superlattices \cite{BelDie82_99} or microwave guides \cite{Kuhl}.
In this work we go one step further and describe a one-dimensional
disordered system, based on the P\"oschl-Teller potential, 
that exhibits a continuum of extended states which is independent of the random or
correlated character of the sequence and of the length of the
system. We then enlarge the family of disordered models showing a perfect
transmission within a continuum energy interval, but this time the total
transparency is provided by the potential itself and it is not due to the
existence of statistical correlations in the disordered sequence. 

\begin{figure}
    \centering
    \includegraphics[width=.7\columnwidth]{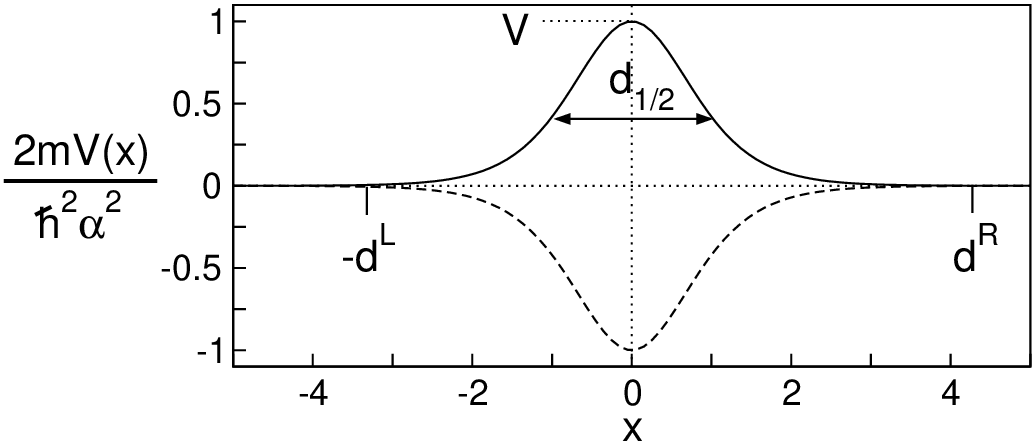}
    \caption{P\"oschl-Teller potential defined in \eqref{eq:PTpot}.}
    \label{fig:PTpot}
\end{figure}

\begin{figure}
    \centering
    \includegraphics[width=.9\columnwidth]{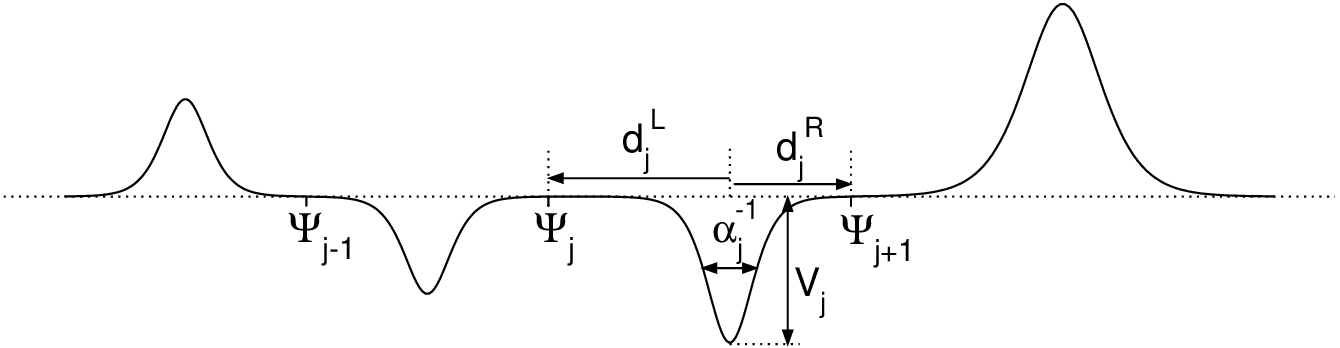}
    \caption{Potential of a disordered P\"oschl-Teller wire}
    \label{fig:cablept}
\end{figure}

Let us consider the general P\"oschl-Teller potential, shown in
Fig.~\ref{fig:PTpot}, and given by
\begin{equation}
    V(x)=\frac{\hslash^2\alpha^2}{2m}\frac{\rmV}{\cosh^2(\alpha x)}.
    \label{eq:PTpot}
\end{equation}
It resembles the form of an atomic well or barrier depending on the sign of
$\rmV$, a dimensionless parameter that together with $\alpha$ determines the
height or depth of the potential. The parameter $\alpha$, with units of
inverse of length, controls the half-width of the potential which reads 
\mbox{$d_{1/2}=2\alpha^{-1}\arc\cosh\sqrt{\smash[b]{2}}$}. The larger
$\alpha$ is the narrower and 
deeper  the potential becomes.
The Schr\"odinger equation for the P\"oschl-Teller potential is analytically solvable and its solutions are well known
\cite{FLUGGE,DabKha21_88}. The 
asymptotic transmission matrix for this potential has been obtained
previously by the authors \cite{CerRod70_04} and it reads
\begin{equation}
    \mathcal{M}=\begin{pmatrix} \rme^{\rmi \varphi} \sqrt{1+w^2} & -\rmi w\\
            \rmi w & \rme^{-\rmi \varphi} \sqrt{1+w^2} \end{pmatrix},
    \label{eq:masinpt}
\end{equation}
where 
\begin{gather}
    w = \frac{\sin(\pi b)}{\sinh(\pi k/\alpha)},\quad  b = \frac{1}{2}+\sqrt{\frac{1}{4}-\rmV},\label{eq:defw}\\
    \varphi =\frac{\pi}{2}+\arg\left\{\frac{\Gamma^2(\rmi k/\alpha)}{\Gamma(b+\rmi
    k/\alpha)\Gamma(1-b+\rmi k/\alpha)}\right\}, \label{eq:defphi}
\end{gather}
$k=\sqrt{2mE}/\hslash$ and $\Gamma(z)$ is the complex Euler gamma function,
also $w$ is always a real quantity as can be seen in its alternative definition 
$w=\cosh(\pi\sqrt{\smash[b]{\rmV-1/4}})/\sinh(k\pi/\alpha)$. The
dimensionless amplitude in terms of $b$ reads
$\rmV=-b(b-1)$ which is the usual form found in the literature. Let us remark that the above expressions are
only valid for positive energies [i.e. $k\in\mathbb{R}$]. From
\eqref{eq:masinpt} the asymptotic probability of transmission is
$T=(1+w^2)^{-1}$. To build a chain with the potentials described, one must do the
approximation of considering that each potential unit has a finite
range. Hence a cut-off  must be included in the P\"oschl-Teller potential.
Using this approximation one obtains matrices suitable to be arranged in
linear chains, applying the composition technique described in Ref. 
\onlinecite{CerRod70_04}. Let us suppose that the potential is appreciable only
inside the interval $[-d^L,d^R]$, as shown in Fig.~\ref{fig:PTpot}.  
Outside this interval the wave function is assumed to be a
superposition of the free particle solutions. Then the transmission matrix
for the cut-off potential reads
\begin{equation}
    \mathbf{M}\!\!=\!\!\begin{pmatrix} \rme^{\rmi\left[
    \varphi+k(d^R+d^L)\right]} \sqrt{1+w^2}  & -\rmi w \rme^{\rmi k(d^R-d^L)}\\
            \rmi w \rme^{-\rmi k(d^R-d^L)} & \rme^{-\rmi \left[\varphi+k(d^R+d^L)\right]} \sqrt{1+w^2} \end{pmatrix}.
    \label{eq:mcutpt}
\end{equation}
The cut-off matrix is the same as the asymptotic one plus an extra phase
term in the diagonal elements that accounts for the total distance $(d^R+d^L)$ during
which the particle feels the effect of the potential, and also an extra
phase term in the off-diagonal elements measuring the 
asymmetry of the cut-off $(d^R-d^L)$. These phases are the
key quantities since they will be responsible for the interference processes
that produce the transmission patterns. In our case due to the rapid decay of the P\"oschl-Teller potential
the cut-off distance admits very reasonable values. In fact we have seen
that for a sensible wide range of the parameters $\alpha$ and $\rmV$, one can
take as a minimum value for the cut-off distance $d_0=2d_{1/2}\simeq
3.5/\alpha$ where $d_{1/2}$ is the half-width. Taking $d^{L,R}\geqslant d_0$ the
connection procedure works really well, as we have checked in all cases
considered by comparing the analytical composition technique versus a
numerical integration of the Schr\"odinger equation for the global
potential. The above matrices can be used to obtain analytical expressions
for the scattering amplitudes of different potential profiles including a
few atoms resembling molecular structures \cite{CerRod70_04}. In this work our main
interest is to consider the transmission matrix \eqref{eq:mcutpt} to
make a continuous  disordered model in the form of a large chain of these
potentials  with random parameters. Let us consider now the
effects of uncorrelated disorder upon this particular model.
From \eqref{eq:mcutpt}
one is led to the following canonical relation among the values of the
electronic states at contiguous sites of the chain,
\begin{equation}
    \Psi_{j+1}=\left(\overline{S}_j+S_{j-1}\frac{K_j}{K_{j-1}}\right)\Psi_j-\frac{K_j}{K_{j-1}}\Psi_{j-1},
    \label{eq:canonicalpt}
\end{equation}
where
\begin{align}
    \overline{S}_j=&
        -w_j\sin\left[k(d^L_j-d^R_j)\right]+\sqrt{\smash[b]{1+w_j^2}} \cos(\Phi_j),\\
    S_j =&\; w_j\sin\left[k(d^L_j-d^R_j)\right]+\sqrt{\smash[b]{1+w_j^2}}
        \cos(\Phi_j), \label{eq:Spt}\\
    K_j =&\; w_j\cos\left[k(d^L_j-d^R_j)\right]+\sqrt{\smash[b]{1+w_j^2}}
        \sin(\Phi_j), \label{eq:Kpt}
\end{align}
in terms of $w$ and $\varphi$ defined in \eqref{eq:defw} and
\eqref{eq:defphi} and $\Phi_j=k(d^L_j+d^R_j)+\varphi_j$. The amplitudes $\Psi_j$ correspond to
the value of the state at the junction points of the potentials as shown in
Fig.~\ref{fig:cablept}, and in this case each potential is determined by
four parameters: $d^L_j,d^R_j,\alpha_j,\rmV_j$.
The form of the canonical relation obtained from the
transmission matrix coincides with the Poincar\'e map derived by S\'anchez and co-workers for 
one-dimensional potentials \cite{SanMac49_94}, in fact expression \eqref{eq:canonicalpt} is
formally independent of the potential model. The canonical equation is essential to obtain the
properties of the disordered system in the thermodynamic limit. From the
canonical relation, the relevant quantities of the disordered composite
P\"oschl-Teller model such as density of states (DOS) and localization length can
be numerically obtained in the thermodynamic limit by using the functional
equation formalism, which has already been successfully applied to other
disordered models by the authors \cite{CerRodEPJB}. The disordered
compositions of P\"oschl-Teller potentials give rise to the emergence of
exciting properties such as fractal DOS, existence of different types of
isolated extended states in the spectrum and the appearance of bound
states for the negative spectrum which can be completely delocalized. A
thorough study of all these features will be reported elsewhere
\cite{future}. This work is devoted to describe the properties of the
disordered system composed of a particular type of P\"oschl-Teller
potentials: the resonant wells. One characteristic feature of the potential
\eqref{eq:PTpot} is that $T=1$ for all energies whenever $b$ is a real integer. 
Hence an absolute resonant transmission occurs for potential wells with $\rmV=-2,-6,-12,-20,\ldots$
independently of the value of $\alpha$. The resonant wells correspond to
potentials with an integer value of $b>1$. 
Since in this case $w=0$, the transmission matrix for a resonant well
becomes diagonal and its non-zero elements are simply the phases
$\rme^{\pm\rmi \Phi_j}$, that is, it is the transmission matrix of a zero
potential. The resonant well for positive energies behaves as a zero
potential with an effective length $L_\textrm{eff}(k)\equiv \Phi_j/k=\varphi/k+(d^R+d^L)$
that depends on the energy.  For a resonant well described by parameters 
$\{d^L_\gamma,d^R_\gamma,\alpha_\gamma,b_\gamma\}$ it can be
proved by induction using the properties of the Gamma function that the
following expression holds,
\begin{equation}
\begin{split}
    \epsilon L_{\textrm{eff}_\gamma}(\epsilon) =&\; \epsilon \frac{\alpha_\gamma(
    d^R_\gamma+d^L_\gamma)}{(\alpha_\gamma/\alpha)}
    -2\sum_{j=1}^{b_\gamma-1}\arc\tan\left(\frac{\epsilon}{j(\alpha_\gamma/\alpha)}\right)\\
    &+(b_\gamma-1)\pi
\label{eq:Leff}
\end{split}
\end{equation}
where $kL_{\textrm{eff}_\gamma}(k)\equiv\epsilon 
L_{\textrm{eff}_\gamma}(\epsilon)$ and the variable $\epsilon\equiv k/\alpha$ is a dimensionless
representation of the energy and $\alpha$ is the reference value for the parameters $\{\alpha_\gamma\}$.
Now let us consider a disordered chain entirely composed of 
resonant wells with different parameters. For positive energies
the functions appearing in the canonical equation of the system
\eqref{eq:canonicalpt} reduce to 
$\overline{S}_j\equiv S_j=\cos\left[kL_{\textrm{eff}_j}(k)\right]$ and
$K_j=\sin\left[kL_{\textrm{eff}_j}(k)\right]$.
It can be easily checked that these functions define the canonical equation for a zero potential 
 where the wave function is evaluated at different distances corresponding to the
effective length of each potential.  It is then clear that the electronic
states for all energies remain extended in the disordered system. The
transmission of the whole structure is maximum for all energies since the
system globally behaves as a zero potential. Let us remark that the fully
resonant behaviour of the P\"oschl-Teller well provided $b_\gamma$ is an integer
is independent of $d^L_\gamma$, $d^R_\gamma$, and $\alpha_\gamma$ as long
as the minimum value for the cut-off distances is preserved. In fact, the real
dimensional depth of the well reads $\hslash^2\alpha^2_\gamma
V_\gamma/(2m)$, hence one can choose at will the depth of the resonant well
by varying $\alpha_\gamma$, although it also means a change in the width of
the potential. Therefore, one can build a disordered chain of resonant wells
with different widths and depths that can even be placed at arbitrary
distances from one another with absolutely no correlations in the sequence,
 which can be completely random indeed, and the structure will behave as a
transparent potential for all energies.  
To our knowledge this is the first theoretical model for which one can build
totally random arrays that exhibit a full continuum of extended states and
hence an interval of complete transparency: the whole positive spectrum.
Let us calculate analytically the distribution of states of these disordered chains in the
thermodynamic limit. For a zero potential of length $L$ the integrated
density of states is trivially 
$\mathcal{N}(k)=Lk/\pi$. From this fact one is led to the conclusion that a
resonant well should provide the spectrum of the system with $k
L_{\textrm{eff}_\gamma}(k)/\pi$ available states with energy less than
$k$. Since all species behave effectively as zero potentials, the IDOS of the
chain per piece of length $\alpha^{-1}$ in the thermodynamic limit is just the composition of the contributions of the
different species with their respective concentrations $\{c_\gamma\}$,
\begin{equation}
    n(\epsilon)=\frac{1}{\pi}\sum_\gamma c_\gamma  \frac{(\alpha_\gamma/\alpha)}{\alpha_\gamma(d^R_\gamma+
    d^L_\gamma)}   \epsilon L_{\textrm{eff}_\gamma}(\epsilon).
\end{equation}
And the DOS would be obtained differentiating with respect to $\epsilon$.
Inserting expression \eqref{eq:Leff} into the latter definition one
finally gets
\begin{equation}
    g(\epsilon)=\frac{1}{\pi}-\frac{2}{\pi}\sum_\gamma c_\gamma
    \frac{(\alpha_\gamma/\alpha)} {\alpha_\gamma(d^R_\gamma+
    d^L_\gamma)} \sum_{j=1}^{b_\gamma-1}
    \frac{j(\alpha_\gamma/\alpha)}{j^2(\alpha_\gamma/\alpha)^2+\epsilon^2}.
    \label{eq:analdos2}
\end{equation}
Using the same reasoning the analytical expression for the DOS can also be
straightforwardly obtained when the 
parameters $\{\alpha_\gamma,d^R_\gamma,d^L_\gamma\}$ obey a continuous distribution.
We have carefully checked how the 
analytical expression reproduces exactly the distribution of states
calculated numerically via the functional equation formalism.
\begin{figure}
    \centering
    \includegraphics[width=.9\columnwidth]{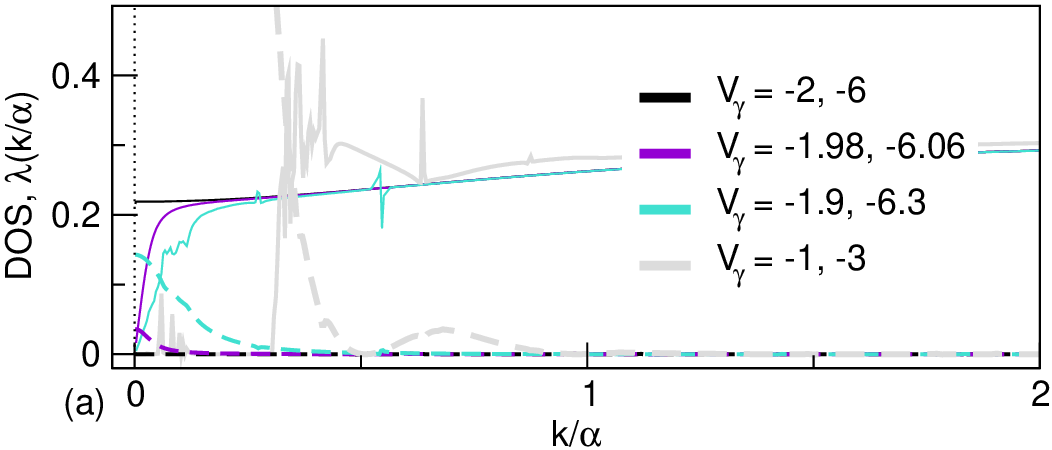}
    \includegraphics[width=.9\columnwidth]{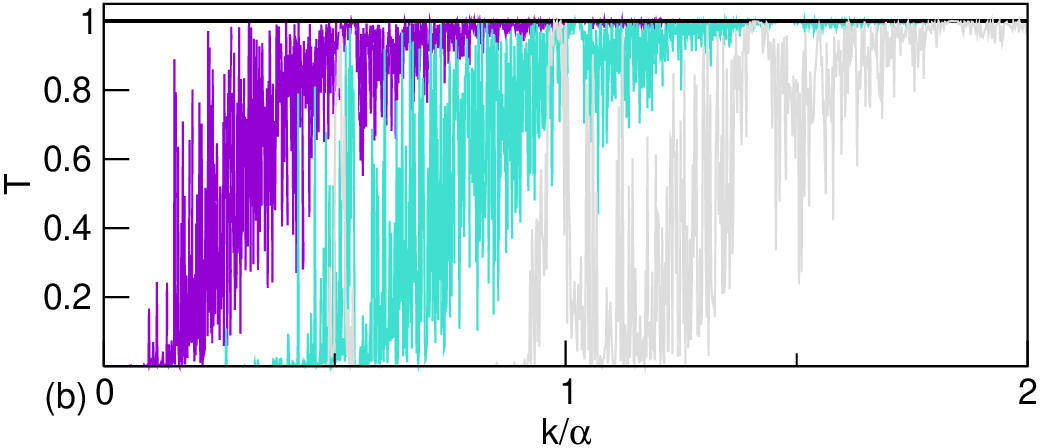}
    \caption{(color online) Tolerance of the properties of a binary resonant chain with their
    parameters. (a) DOS(solid line) and Lyapunov exponent(dashed line) in the thermodynamic
    limit. (b) Transmission patterns for a $1000$-atom random
    sequence. Cases have been considered where both 
    dimensionless amplitudes are deviated $1\%$, $5\%$ and $50\%$ from the
    resonant values $\rmV_1=-2$, $\rmV_2=-6$. $\alpha_\gamma=\alpha$ and
    $d^L_\gamma=d^R_\gamma=4/\alpha$ for both species.}
    \label{fig:tolerance}
\end{figure}
The DOS for the resonant chains is a continuous and smooth function without
gaps that does not vanish for zero energy, and it registers relatively small changes by varying the concentrations or the
number of different resonant wells.
In Fig.~\ref{fig:tolerance} the tolerance of the properties of a binary
resonant chain are evaluated when their parameters are deviated from the resonant
values. As can be seen, a small change of the parameters mean the loss of the
full resonant behaviour for all energies. Nevertheless for deviations of
order $1\%-5\%$ in the dimensionless amplitudes, the efficiency of transmission is still much
higher than for any other non-resonant binary chain composed of wells.
Naturally, for the resonant chains the Lyapunov exponent $\lambda$ in the
thermodynamic limit, corresponding to the inverse of the localization length, calculated via the functional
equation vanishes for all energies. It can also be checked that the inverse
participation ratio for finite resonant chains as a function of the energy is simply a straight line at the value
$N^{-1}$ where $N$ is the number of potentials, as it must be for flat
extended states. 

\begin{figure}
    \centering
    \includegraphics[width=.95\columnwidth]{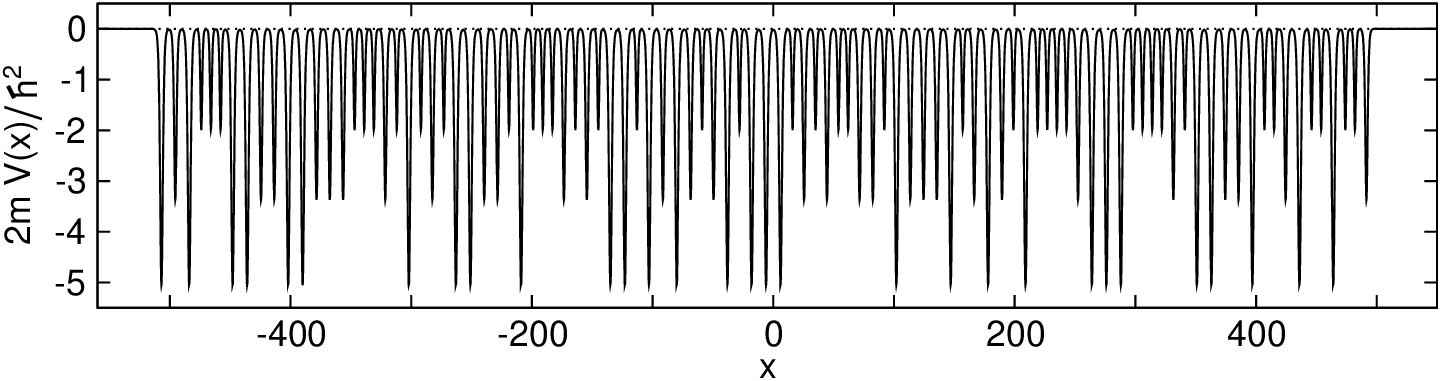}\\[2mm]
    \includegraphics[width=.9\columnwidth]{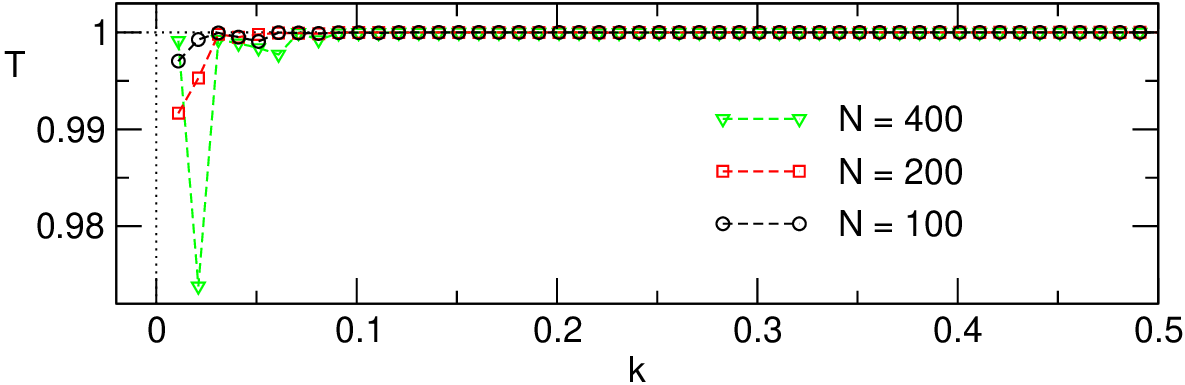}
    \caption{(color online) Transmission probability for random resonant chains of
    P\"oschl-Teller wells, calculated by solving numerically the Schr\"odinger
    equation for the continuum spectrum. The upper box shows the random 
    potential profile for $100$ potentials. For all lengths the chains 
    include three different species with symmetric cut-off $(d^L_\gamma=d^R_\gamma=d_\gamma)$. The parameters 
    are $\{\alpha_\gamma,\rmV_\gamma,d_\gamma\}[c_\gamma]$: $\{1,-2,4\}[0.4]$,
    $\{0.75,-6,5.5\}[0.3]$, $\{0.65,-12,6\}]0.3]$.}
    \label{fig:resoerror}
\end{figure}

One must not forget that the transmission matrix proposed
for the P\"oschl-Teller potential is an approximation, since we have
assumed that at the cut-off distance the asymptotic form of the
states can be used. In fact, this approximation is quite correct; the
error that it entails is almost irrelevant for an individual
potential and the larger the cut-off distance is the smaller the error
becomes. However it might happen that when applying the composition
procedure of the potentials to build a disordered array, these small
individual deviations give rise to an error growing exponentially with
the length of the chain. If it were true, then the behaviour of a real 
continuous composition of P\"oschl-Teller units [i.e. the sum of all the
contributions of the potentials centred at different positions] would be
far from the results obtained using our techniques. In particular it would
be dramatic for a resonant chain for which its resonant behaviour and the
delocalization of the electronic states could disappear in the real
continuous composition. To show that this exponential error does not occur,
we have calculated the transmission probability of several random
resonant chains with $100$, $200$ and $400$ potentials, by integrating numerically
the Schr\"odinger equation, via a spatial discretization of the whole continuous potential of the
chain, that is taking into account the superposition of all potential wells centred at
their respective positions. 
In Fig.~\ref{fig:resoerror} it can be
observed how for very low energies ($k<0.05$) a small deviation appears
from $T=1$, that for the longest chain is less than $3\cdot
10^{-2}$. Although this deviation seems to increase slightly with the length of the chain,
its effect does not noticeably distort the resonant behaviour of the chain. Then,  
 our composition procedure describes faithfully the
properties of the real continuous composite potential profile.

In summary we have described a class of random resonant chains with a continuum of delocalized states. The
composition of resonant P\"oschl-Teller wells behaves as a transparent
potential for all positive energies. As long as the dimensionless amplitude of the well
belongs to an infinite set of discrete values that provide the resonant
behaviour, the rest of the parameters of the well can be varied randomly, 
therefore the configuration of the resonant chain is quite versatile. And
of course the delocalization of the electronic states for positive energies
is absolutely independent of the random or correlated character of the
disordered sequence. Then, at least it is possible to find a theoretical
model for which disordered arrays of potentials exhibits a full continuum
of extended states which is independent of the length of the system. It is
in principle a pure academic model whose properties are tightly bound to the
functional dependence of the potentials. Hence, its real importance depends up to a
point on the possibility to reproduce experimentally such a
structure. Semiconductor heterostructures may be considered as 
applicants for this task. Advances in the epitaxial growing techniques have
made it possible to manipulate the profiles of the band conduction inside the
heterostructure in order to build for example confining parabolic wells. Then if
not now, perhaps in the future it might be possible to control the growing process of 
semiconductor samples in such a manner that the spatial profile of the band
conduction follows the functional dependence of the P\"oschl-Teller well
and therefore having the possibility to check experimentally the predicted behaviour.
The P\"oschl-Teller potential shows an ensemble of very interesting
properties which will be described in detail \cite{future} and also a
completely  new behaviour not expected from a
disordered system.

We thank E. Diez for useful discussions. We acknowledge financial support
from DGICYT under contract No. BFM2002-02609.

\end{document}